# Single-domain ferromagnet of noncentrosymmetric uniaxial magnetic ions and magnetoelectric interaction

Hyun-Joo Koo[a], Elijah E. Gordon[b] and Myung-Hwan Whangbo*,[b],[c]

[a] Department of Chemistry and Research Institute for Basic Sciences Kyung Hee University, Seoul 02447, Republic of Korea

[b] Department of Chemistry, North Carolina State University, Raleigh, NC 27695-8204, USA

   Email: mike_whangbo@ncsu.edu

[c] Group SDeng, State Key Laboratory of Structural Chemistry, Fujian Institute of Research on the Structure of Matter (FJIRSM), Chinese Academy of Sciences (CAS), Fuzhou 350002, China

Abstract: The feasibility of a single-domain ferromagnet based on uniaxial magnetic ions was examined. For a noncentrosymmetric uniaxial magnetic ion of magnetic moment μ at a site of local electric dipole moment p, it is unknown to date whether μ prefers to be parallel or antiparallel to μ. The nature of this magnetoelectric interaction was probed in terms of analogical reasoning based on the Rashba effect and density functional calculations. We show that μ and p prefer an antiparallel arrangement, predict that Fe-doped CaZnOS is a single-domain ferromagnet like a bar magnet, and find the probable cause for the ferromagnetism and weak magnetization hysteresis in Fe-doped hexagonal ZnO and ZnS at very low dopant concentrations.

    A ferromagnetic (FM) material exhibits magnetization hysteresis because it forms numerous FM domains of different moment directions typically to diminish the magnetic dipole-dipole (MDD) interactions.[1] The MDD interactions, being long-range interactions, make it energetically unfavorable for a ferromagnet to have a single FM domain. When a ferromagnet is sufficiently small in size so that its MDD interactions are negligible, it can have a single FM domain and becomes a superparamagnet.[2] It is interesting to think about the feasibility of a single-domain ferromagnet (SDF) even when its size is not small. In most cases, ferromagnetism arises when spin exchange interactions between adjacent magnetic ions are FM. Since spin exchange interactions are short-range interactions, such a ferromagnetism requires the presence of magnetic ions in proximity, which in turn makes MDD interactions substantial. When magnetic ions are well separated, both spin exchange and MDD interactions can be made negligible. In this case, ferromagnetism is possible if it is based on uniaxial magnetism.[3] Uniaxial magnetic ions have a nonzero magnetic moment μ only in one direction (by convention, the z-direction) in

space. If such ions are arranged with their z-axes aligned in one common direction and enough separation between them, a SDF should be feasible regardless of its size.

For a uniaxial ion of moment $\mu$ at a site of inversion symmetry (i.e., a centrosymmetric uniaxial magnetic ion), the moment orientation along the ||z direction is identical in energy to that along the -||z direction. So far, there has been no clear understanding as to whether this is also true for a uniaxial magnetic ion at a site of no inversion symmetry (i.e., a noncentrosymmetric uniaxial magnetic ion), which is characterized by a nonzero local electric dipole moment $\mathbf{p}$ unless the site has additional symmetry to make the ||z and -||z directions equivalent. In this Communication, we probe this question on the basis of analogical reasoning and DFT calculations to find that the magnetoelectric interaction between $\mu$ and $\mathbf{p}$ can be described by the energy $\alpha\mu\cdot\mathbf{p}$ with positive constant $\alpha$, predict that Fe-doped CaZnOS is a SDF, and explain why Fe-doped hexagonal ZnO and ZnS exhibit ferromagnetism and weak magnetization hysteresis at very low dopant concentrations.

For a magnetic ion at a coordination site of $C_n$ ($n \geq 3$) rotational symmetry, its d-states are split to have doubly-degenerate sets, {xz, yz} and {xy, $x^2$-$y^2$}, with the z-axis taken along the rotational axis.[3d, 3e] A uniaxial magnetic ion has the electron configuration that includes an unevenly-filled degenerate level, e.g., (xz, yz)$^1$, (xz, yz)$^3$, (xy, $x^2$-$y^2$)$^1$ or (xy, $x^2$-$y^2$)$^3$, which gives rise to unquenched orbital momentum $\mathbf{L}$. Several coordinate environments of known uniaxial magnetic ions are presented in **Fig. 1**. For convenience, the $3z^2$-$r^2$ orbital will be referred to as the 1a level, and the lower and higher energy sets of

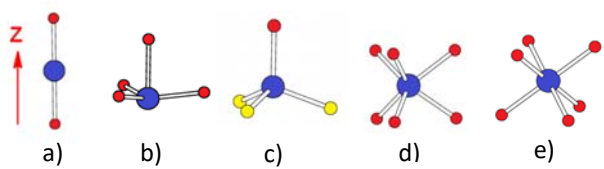

Figure 1. $ML_n$ polyhedra (n = 2, 4, 6) containing uniaxial magnetic ions M, where cobalt spheres = M, and small spheres = L. The z-axis taken along the $C_3$ rotational axis in b) – e).

{xz, yz} and {xy, $x^2$-$y^2$} as the 1e and 2e levels, respectively. The uniaxial electron configuration (1e)$^3$(2e)$^2$(1a)$^1$ (**Fig. 2a**) has been used to describe the Fe$^{2+}$ (S = 2, d$^6$) ions in the FeC$_2$ dumbbells (**Fig. 1a**) of Fe{C(SiMe$_3$)$_3$}$_2$ [3a, 4] and the FeN$_4$ trigonal pyramids (**Fig. 1b**) of [K(solvent)$_n$][(tpa$^{Mes}$)Fe],[5] where tpa = tris(pyrrolyl-$\alpha$-methyl) amine. It can also be used for the Fe$^{2+}$ (S = 2, d$^6$) ions of the FeOS$_3$ tetrahedra (**Fig. 1c**) in the layered phase CaOFeS (see below).[6] The uniaxial configuration (1a)$^2$(1e)$^3$(2e)$^2$ (**Fig. 2b**) has been employed to discuss the Co$^{2+}$ (S = 3/2, d$^7$) ions in the CoO$_6$ trigonal prisms (**Fig. 1d**) of Ca$_3$CoMnO$_6$[3b, 3d, 7] and the Fe$^+$ (S = 3/2, d$^7$) ions in the FeC$_2$ dumbbells (**Fig. 1a**) of [K(crypt-222)][Fe{C(SiMe$_3$)$_3$}$_2$].[8] Uniaxial magnetic ions are found at octahedral sites as well (**Fig. 1e**). Provided that the z-axis is taken along one $C_3$ rotational axis of an octahedron, the $t_{2g}$ state is described by 1a and 1e', and the $e_g$ state by 2e' level.[3d, 3e] The high-spin Fe$^{2+}$ (S = 2, d$^6$) ion at an octahedral site of BaFe$_2$(PO$_4$)$_2$ [9] has the uniaxial configuration (1a)$^2$(1e')$^3$(2e')$^2$ (**Fig. 2c**), and the low-spin Ir$^{4+}$ (S = 1/2, d$^5$) at an octahedral site of Sr$_3$NiIrO$_6$ the uniaxial configuration (1a)$^2$(1e')$^3$ (**Fig. 2d**).[3f, 10]

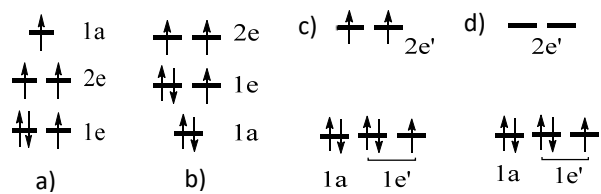

Figure 2. Electronic configurations leading to uniaxial magnetism. In c) and d) for an octahedron, the z-axis is taken along one of its $C_3$ rotational axis. Then, the xz and xy orbitals mix, and so do the $x^2-y^2$ and yz orbitals. These orbitals mixings give rise to the two sets of degenerate levels 1e' and 2e'.[3d, 3e]

Consider a noncentrosymmetric uniaxial magnetic ion of magnetic moment **μ** located at a site of local electric dipole moment **p** along the rotational axis. Our question is whether the moment orientations parallel and antiparallel orientations to **p** (hereafter, ‖**p** and -‖**p**, respectively) are identical in energy. So far, none is known about the nature of the magnetoelectric interaction between **μ** and **p**, although the free-energy expansion [11] with respect to the electric field **E** and the magnetic field **H** leads to the magnetoelectric term $\alpha_{ij}E_iH_j$, where $E_i$ and $H_j$ are the Cartesian components of **E** and **H**, respectively, with $\alpha_{ij}$ as the associated constant. To gain insight into the magnetoelectric interaction, we resort to an analogical reasoning based on the Rashba effect.[12] This effect deals with a nonmagnetic metal or a nonmagnetic semiconductor, for which the up-spin and down-spin subbands of any given band are degenerate. If such a system lacks inversion symmetry and consists of heavy elements with strong spin-orbit coupling (SOC), the subbands of a given band become nondegenerate due to a combined effect of SOC and inversion-symmetry loss.[12] For a magnetic ion of spin momentum **S**, orbital momentum **L** and SOC constant λ, the associated SOC is described by the term λ**S·L**, and the energy split between the up-spin and down-spin subbands increases with increasing the strength of SOC.

The Rashba effect deals with a situation when the split subbands are either both filled or both empty. Here we consider a half-filled band for which only the lower-energy one of the split subbands is filled. In general, the energetic behavior of a discrete molecule is well described by that of a hypothetical solid containing the molecule in each unit cell with large repeat distances. Imagine such a solid of a molecular species containing a noncentrosymmetric uniaxial magnetic ion in which the electric dipole moment **p** of each molecule is pointed in one common direction, so the solid is noncentrosymmetric. For simplicity, each magnetic ion may be assumed to have one orbital and one electron. In the absence of the Rashba effect, this solid has one flat band with energy at the orbital level of the magnetic ion for all wave vectors k, and the up-spin and down-spin subbands are degenerate. Given no inversion symmetry and unquenched orbital momentum **L** at each ion site, the Rashba effect takes place and hence splits the up-spin and down-spin subbands. With one electron per site, only the lower-energy subband becomes occupied so that each magnetic ion has one identical spin state filled, with the other spin state empty. In other words, for an isolated noncentrosymmetric uniaxial magnetic ion, one moment orientation should be lower in energy than its opposite orientation, i.e., the moment orientations along the ‖**p** and -‖**p** directions should be different in energy.

The layered phase CaOFeS consists of hexagonal FeOS layers made up of sulfur-corner-sharing $FeOS_3$ tetrahedra (**Fig. 3a**), in which all Fe-O bonds are oriented in one direction perpendicular to the layer. These layers are stacked along the c-direction (**Fig. 3b**) such that there occur two FeOS layers per unit cell.[6] The $Fe^{2+}$ ions of the FeOS layers form hexagonal lattices (**Fig. 3c**). As already mentioned, the $Fe^{2+}$ ion of each $FeOS_3$ tetrahedron is uniaxial. The magnetic structure of CaOFeS determined from powder neutron diffraction measurements shows that each hexagonal lattice of CaOFeS has the antiferromagnetic (AFM) arrangement depicted in **Fig. 3d** with spin at each site aligned along the Fe-O bond, and adjacent hexagonal lattices are antiferromagnetically coupled. Thus, in the ordered magnetic structure of CaOFeS, half the spins have their spin vectors aligned along the Fe→O bonds [i.e., ||(Fe→O)], and the remaining ones along the O→Fe bonds [i.e., -||(Fe→O)]. This spin arrangement is unusual from the viewpoint of spin frustration expected for each hexagonal layer. The intralayer exchange $J_1$ as well as the interlayer exchanges $J_2$ and $J_3$ (**Fig. 3c**) evaluated by performing energy mapping analysis [3c] based on DFT+U calculations (see Experimental Section) show that $J_1$ – $J_3$ values are all AFM, and $J_1$ dominates over $J_2$ and $J_3$ (**Table S1**). Thus each hexagonal spin lattice is spin-frustrated and is expected to have a compromised 120° noncollinear spin arrangement to reduce the extent of spin frustration.[13] However, this noncollinear arrangement requires the $Fe^{2+}$ ions to adopt a spin orientation away from their local z-axes. The latter is prevented because each $Fe^{2+}$ ion has uniaxial magnetism.

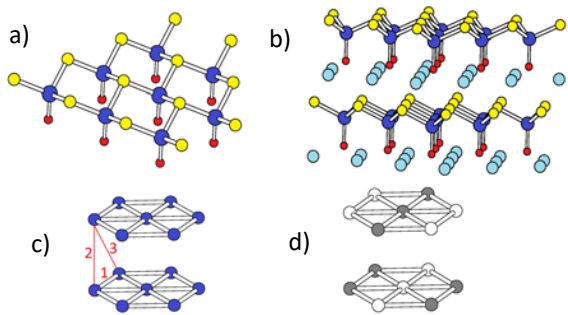

Figure 3. a) FeOS layer made up of sulfur-corner-sharing $FeOS_3$ tetrahedra, where yellow sphere = S, red sphere = O, and cobalt sphere = Fe. b) Stacking of adjacent FeOS layers in CaOFeS, where cyan sphere = Ca. c) Hexagonal lattices of $Fe^{2+}$ ions, where the numbers 1 – 3 represent the spin exchange paths $J_1$ – $J_3$, respectively. d) Ordered AFM state of CaOFeS, where the shaded and unshaded circles represent up-spin and down-spin $Fe^{2+}$ sites, respectively.

To determine which direction, ||(Fe→O) or -||(Fe→O), is energetically more favorable for the $Fe^{2+}$ spin of each $FeOS_3$ tetrahedron, we need to construct a well-separated arrangement of $FeOS_3$ tetrahedra so that there is no spin exchange interaction between $Fe^{2+}$ ions. CaZnOS [14] is isostructural with CaOFeS, and consists of nonmagnetic $Zn^{2+}$ ions in the $ZnOS_3$ tetrahedra. In recent years, CaZnOS has been actively studied for the long-lasting luminescence properties of its doped phases, and CaZnOS:Mn and CaZnOS:Cu are known.[15] Thus, it should be possible to prepare CaZnOS:Fe samples, in which $FeOS_3$ tetrahedra are well separated from each other. To simulate such samples, we construct a (2a, 2b, c) supercell of CaZnOS, which has eight $Zn^{2+}$ sites per supercell, and replace only one $Zn^{2+}$ ion with a $Fe^{2+}$ ion per supercell. The resulting CaZnOS:Fe structure will be referred to as the (2×2×1)-Fe model. For this model, we consider the

state with the Fe$^{2+}$ spins along the ||(Fe→O) direction and that along the -||(Fe→O) direction. We determine the relative energies of these two spin orientations for the (2×2×1)-Fe model by performing DFT+U+SOC calculations (see Experimental Section). The results of these calculations (**Table S2**) reveal that the moment orientation along ||(Fe→O) is more stable than that along -||(Fe→O) direction by ~2 meV/Fe. Then, for a uniaxial magnetic ion of moment **μ** at a site of no inversion symmetry with local electric dipole moment **p**, the energy E of the magnetoelectric interaction between **μ** and **p** can be written as

$$E = \alpha \mathbf{\mu} \cdot \mathbf{p} \qquad (1)$$

with material-dependent constant $\alpha$. The dipole moment of the Fe-O bond has the O$^{\delta-}$→Fe$^{\delta+}$ direction, which is antiparallel to the preferred magnetic moment direction. If the local electric dipole moment of the FeOS$_3$ tetrahedron is dominated by that of the Fe-O bond, then $\alpha > 0$. We discuss this point further by considering the single-ion magnet, [K(solvent)$_n$][(tpa$^{Mes}$)Fe],[5] in which the FeN$_4$ trigonal pyramid (**Fig. 1b**) has one apical Fe-N$_a$ and three basal Fe-N$_b$ bonds. The electric dipole moment of the FeN$_4$ trigonal prism will be dominated by the dipole moment of the Fe-N$_a$ bond, because those of the three Fe-N$_b$ bonds will be nearly canceled out by their geometrical arrangement. Our electronic band structure calculations using the simplified molecule [(tpa$^H$)Fe], which results when the mesityl group is replaced with H, show that the magnetic moment orientation along ||(Fe→N$_a$) is more stable than that along -||(Fe→N$_a$) direction by ~6 meV/Fe (**Table S3**). The dipole moment of the Fe-N$_a$ bond has the N$^{\delta-}$→Fe$^{\delta+}$ direction, which is antiparallel to the preferred moment direction so that $\alpha > 0$, as speculated for the FeOS$_3$ tetrahedron.

CaZnOS:Fe would be a SDF in which all the Fe$^{2+}$ spins are aligned along the ||(Fe→O) direction. Consequently, it would be a bar magnet, for which one spin alignment is a minimum-energy state, while the opposite spin alignment is not. The smallest bar magnet would be a single-ion magnet containing an noncentrosymmetric uniaxial magnetic ion as found for [K(solvent)$_n$][(tpa$^{Mes}$)Fe]. In Fe-doped hexagonal ZnO and ZnS, consisting of corner-sharing ZnL$_4$ (L = O, S) tetrahedra, the doped Fe$^{2+}$ ions are uniaxial. The FeL$_4$ tetrahedra should be nearly regular in shape, so the moment orientations pointing to the four C$_3$ axes (i.e., the Zn-L bonds) would be similar in energy. Thus, under a magnetic field, all Fe$^{2+}$ moments of ZnL:Fe can easily line up with the field. Nevertheless, if the FeL$_4$ tetrahedra have a weak distortion from the local T$_d$ to a local C$_3$ symmetry, then the moment orientation toward a particular C$_3$ axis would be slightly preferred. To reduce the lattice strain, a given ZnL:Fe sample would form numerous domains of different C$_3$-axis (hence moment) orientations. These are the most probable reasons for why ZnL:Fe samples exhibit ferromagnetism and weak magnetization hysteresis when the doping level is so low that the MDD and spin exchange interactions among the dopants are negligible.[20, 21]

In summary, for a noncentrosymmetric uniaxial magnetic ion, the magnetoelectric interaction can be described by the energy $\alpha \mathbf{\mu} \cdot \mathbf{p}$ with coefficient $\alpha > 0$ so that **μ** and **p** prefer to be antiparallel to each other. Fe-doped CaZnOS is predicted to be a SDF like a bar magnet. The ferromagnetism and weak magnetization hysteresis observed for Fe-doped ZnO and ZnS originate from the uniaxial magnetic ions Fe$^{2+}$.

**Experimental Section**

Our spin-polarized DFT calculations employed the frozen-core projector augmented wave method[16] encoded in the Vienna ab initio simulation package,[17] and the generalized-gradient approximation of Perdew, Burke and Ernzerhof[18] for the exchange-correlation functional. The electron correlation in Fe 3d states was taken into consideration in terms of the DFT+U method[19] by adding the effective on-site repulsion $U_{eff}$ of 4 and 5 eV on the Fe sites. The spin exchange interactions of CaOFeS were evaluated by using the energy-mapping analysis[3c] based DFT+U calculations. Spin-polarized DFT+U+SOC calculations were employed to determine the preferred moment directions of the (2×2×1)-Fe model of CaZnOS:Fe and a hypothetical solid containing one [(tpa$^H$)Fe]$^-$ anion and one [K(solvent)$_n$]$^+$ cation per unit cell. Other computational details are given in the supporting information.

**Acknowledgements**

The work used resources of the HPC at NCSU as well as the NERSC Centre, a DOE Office of Science User Facility supported by the Office of Science of the U.S. Department of Energy under Contract No. DE-AC02-05CH11231.

**Keywords:** Spin-orbit coupling • Uniaxial magnetic ion • Loss of inversion symmetry • Magnetoelectric interaction

[1] N. W. Ashcroft and N. D. Mermin, *Solid State Physics*, Saunders College, Philadelphia, 1976, pp. 718–722.
[2] M. Knobel, W. C. Nunes, L. M. Socolovsky, E. De Biasi, J. M. Vargas, J. C. Denardin, *J. Nanosci. Nanotechnol.* **2008**, *8*, 2836-2857.
[3] a) D. Dai, M.-H. Whangbo, *Inorg. Chem.* **2005**, *44*, 4407-4414. b) D. Dai, H. J. Xiang, M.-H. Whangbo, *J. Comput. Chem.* **2008**, *29*, 2187-2209. c) H. J. Xiang, C. Lee, H.-J. Koo, X. G. Gong, M.-H. Whangbo, *Dalton Trans.* **2013**, *42*, 823-853. d) M.-H. Whangbo, E. E. Gordon, H. J. Xiang, H.-J. Koo, C. Lee, *Acc. Chem. Res.* 2015, *48*, 3080-3087. e) E. E. Gordon, H. J. Xiang, J. Köhler, M.-H. Whangbo, *J. Chem. Phys.* **2016**, *144*, 114706.
[4] W. M. Reiff, A. M. LaPointe, H. Witten, *J. Am. Chem. Soc.* **2004**, *126*, 10206-10207.
[5] W. H. Harman, T. D. Harris, D. E. Freedman, H. Fong, A. Chang, J. D. Rinehart, A. Ozarowski, M. T. Sougrati, F. Grandjean, G. J. Long, J. R. Long, C. J. Chang, *J. Am. Chem. Soc.* **2010**, *132*, 18115-18126.
[6] S. F. Jin, Q. Huang, Z. P. Lin, Z. L. Li, X. Z. Wu, T. P. Ying, G. Wang, X. L. Chen, *Phys. Rev. B* **2015**, *91*, 094420.
[7] Y. Zhang, H. J. Xiang, M.-H. Whangbo, *Phys. Rev. B* **2009**, *79*, 054432.
[8] J. M. Zadrozny, D. J. Xiao, M. Atanasov, G. J. Long, F. Grandjean, F. Neese, J. R. Long, *Nature Chem.* **2013**, *5*, 577-581.
[9] H. Kabbour, R. David, A. Pautrat, H-J. Koo, M.-H. Whangbo, G. André, O. Mentré, *Angew. Chem. Int. Ed.* **2012**, *51*, 11745-11749.
[10] a) S. Toth, W. Wu, D. T. Adroja, S. Rayaprol, E. V. Sampathkumaran, *Phys. Rev. B.* **2016**, *93*, 174422. b) E. Lefrancois, L. C. Chapon, V. Simonet, P. Lejay, D. Khalyavin, S. Rayaprol, E. V. Sampathkumaran, R. Ballou, D. T. Adroja, *Phys. Rev. B,* **2014**, *90*, 014408.
[11] K. F. Wang, J.-M. Liu, Z. F. Ren, *Adv. Phys.* **2009**, *58*, 321-448.

[12] a) E. I. Rashba, *Sov. Phys. Solid State* **1960**, *2*, 1109. b) A. Manchon, H. C. Koo, J. Nitta, S. M. Frolov, R. A. Duine, *Nature Mater.* **2015**, *14*, 871-882.

[13] a) J. E. Greedan, *J. Mater. Chem.* **2001**, *11*, 37-53. b) D. Dai, M.-H. Whangbo, *J. Chem. Phys.* **2004**, *121*, 672.

[14] T. Sambrook, C. F. Smura, S. J. Clarke, K. M. Ok, P. S. Halasyamani, *Inorg. Chem.* **2007**, *46*, 2571-2574.

[15] a) D. Tu. D. Peng, C.-N. Xu, A. Yoshida, *J. Cer. Soc. Jpn.* **2016**, *124*. 702-705. b) J.-C. Zhang, L.-Z. Zhao, Y.-Z. Long, H.-D. Zhang, B. Sun, W.-P. Han, X. Yan, X. Wang, *Chem. Mater.* **2015**, *27*, 7481-7489. c) D. Tu, C.-N. Xu, Y. Fujio, S. Kamimura, Y. Sakada, N. Ueno, *Appl. Phys. Lett.* **2016**, *105*, 011908.

[16] a) P. E. Blöchl, *Phys. Rev. B* **1994**, *50*, 17953-17979. b) G. Kresse, D. Joubert, *Phys. Rev. B* **1999**, *59*, 1758-1775.

[17] a) G. Kresse, J. Furthmüller, *Phys. Rev. B* **1996**, *54*, 11169-11186. b) G. Kresse, J. Furthmüller, *Comput. Mater. Sci.* **1996**, *6*, 15-50.

[18] J. P. Perdew, K. Burke, M. Ernzerhof, *Phys. Rev. Lett.* **1996**, *77*, 3865- 3868.

[19] S. L. Dudarev, G. A. Botton, S. Y. Savrasov, C. J. Humphreys, A. P. Sutton, *Phys. Rev. B* **1998**, *57*, 1505.

[20] a) W. Cheng, X. Ma, *J. Phys.: Conf. Ser.* **2009**, *152*, 01239. b) D. Karmarkar, S. K. Mandal, R. M. Kadam, S. L. Paulose, A. K. Rajarajan, T. K. Nath, A. K. Das, I. Dasgupta, G. P. Das, *Phys. Rev. B* **2007**, *75*, 144404

[21] a) E. Nie, D. Liu, Y. Zhang, X. Bai, L. Yi, Y. Jin, Z. Jiao, X. Sun, *Appl. Surf. Sci.* **2011**, *257*, 8762-8766. b) S. Kumar, N. K. Verma, *J. Electr. Mater.* **2015**, *44*, 2829-2834.

COMMUNICATION

We explored the feasibility of a single-domain ferromagnet of uniaxial magnetic ions, the nature of the magnetoelectric interaction for a noncentrosymmetric uniaxial magnetic ion, and the cause for ferromagnetism and weak magnetization hysteresis in Fe-doped hexagonal ZnO and ZnS.

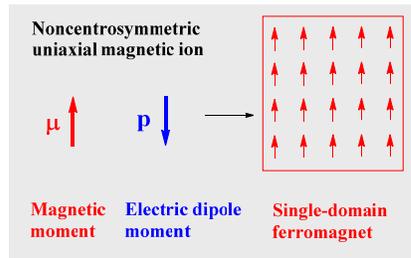

H.-J. Koo, E. E. Gordon, M.-H. Whangbo*

*Page No. – Page No.*

**Single-domain ferromagnet of noncentrosymmetric uniaxial magnetic ions and magnetoelectric interaction**

# Single-domain ferromagnet of noncentrosymmetric uniaxial magnetic ions and magnetoelectric interaction

Hyun-Joo Koo[a], Elijah E. Gordon[b] and Myung-Hwan Whangbo*,[b],[c]

**Abstract:** The feasibility of a single-domain ferromagnet based on uniaxial magnetic ions was examined. For a noncentrosymmetric uniaxial magnetic ion of magnetic moment **μ** at a site of local electric dipole moment **p**, it is unknown to date whether **μ** prefers to be parallel or antiparallel to **μ**. The nature of this magnetoelectric interaction was probed in terms of analogical reasoning based on the Rashba effect and density functional calculations. We show that **μ** and **p** prefer an antiparallel arrangement, predict that Fe-doped CaZnOS is a single-domain ferromagnet like a bar magnet, and find the probable cause for the ferromagnetism and weak magnetization hysteresis in Fe-doped hexagonal ZnO and ZnS at very low dopant concentrations.

**Table of Contents**

[1] Spin exchange constants $J_1 - J_3$ of CaOFeS

[2] Relative energies of the states of CaOZnS:Fe in which the $Fe^{2+}$ moments are oriented along the ||(Fe→O) and -||(Fe→O) directions

[3] Relative energies of the two states of [K(1,2-dimethoxyethane)$_4$][(tpa$^{Mes}$)Fe] in which the $Fe^{2+}$ moments are oriented along the ||(Fe→N$_a$) and -||(Fe→ N$_a$) directions

[4] Author contributions

## [1] Spin exchange constants $J_1 - J_3$ of CaOFeS

1) Spin exchange paths $J_1 - J_3$ of CaOFeS

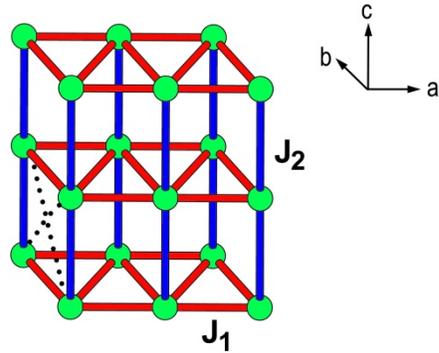

Figure 1. Spin exchange paths $J_1 - J_3$. The red, blue and black dotted lines represent $J_1$, $J_2$ and $J_3$, respectively.

2) Ordered spin states used to evaluate $J_1 - J_3$.

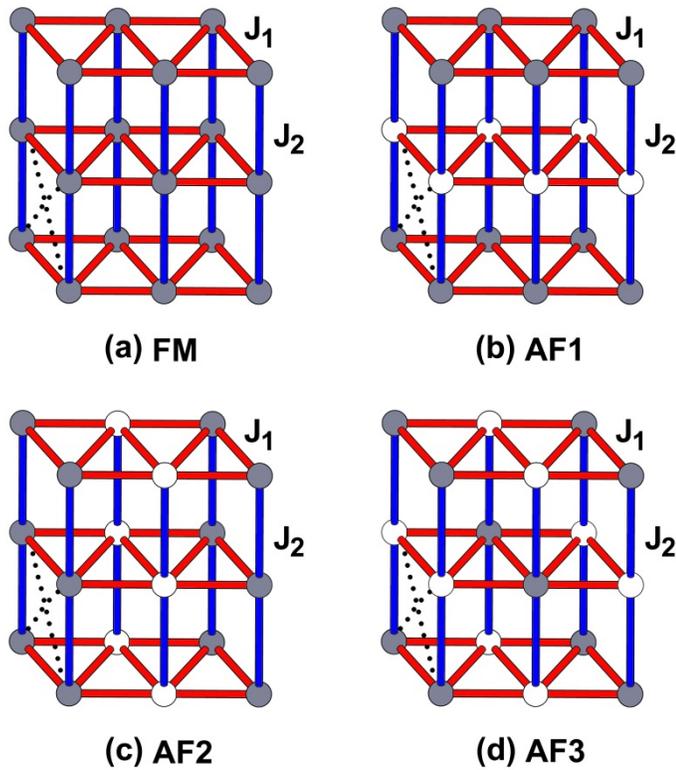

Figure 2. Ordered spin states FM, AF1, AF2, and AF3 of CaOFeS.

3) The energies of the ordered spin states per (2a, b, c) supercell in terms of the spin exchange constants $J_1 - J_3$.

$$FM = (-12J_1 - 4J_2 - 24J_3)(N^2/4)$$

$$AF1 = (-12J_1 + 4J_2 + 24J_3)(N^2/4)$$

$$AF2 = (+4J_1 - 4J_2 + 8J_3)(N^2/4)$$

$$AF3 = (+4J_1 + 4J_2 - 8J_3)(N^2/4)$$

where N = 4, i.e., the number of unpaired spins in the high-spin $Fe^{2+}$ ion.

4) The spin exchange contants in terms of the energies of the ordered spin states.

$$J_3 = (1/64)(4/N^2)[(AF1 - FM) - (AF3 - AF2)]$$

$$J_1 = (1/32)(4/N^2)[(AF2 - AF1) - (FM - AF3)]$$

$$J_2 = (1/8)[(AF3 - AF2)(4/N^2) + 16J_3]$$

5) The values of $J_1 - J_3$ determined from DFT+U calculations for the ordered spin states FM, AF1, AF2, and AF3 of CaOFeS using $U_{eff}$ = 4 and 5 eV on the Fe atoms.

Table S1. Spin exchange constants (in $k_B K$) obtained from DFT+U calculations for CaOFeS.

|     | $U_{eff}$ = 4 eV | $U_{eff}$ = 5 eV |
| --- | --- | --- |
| $J_1$ | -25.81 | -19.97 |
| $J_2$ | -1.83 | -1.46 |
| $J_3$ | -0.55 | -0.44 |

**[2] Relative energies of the states of CaOZnS:Fe in which the Fe$^{2+}$ moments are oriented along the ||(Fe→O) and -||(Fe→O) directions**

The relative energies $\Delta E$ (meV/Fe) of these two states were calculated by performing DFT+U+SOC calculations for the (2×2×1)-Fe model (see the text) with $U_{eff}$ = 4 and 5 eV with a set of (5×5×3) k-points, the SCF convergence criterion of 10$^{-7}$ eV, and the planewave cutoff energy of 450 eV.

Table S2. Relative energies $\Delta E$ (meV/Fe)

|  | $U_{eff}$ = 4 eV | $U_{eff}$ = 5 eV |
| --- | --- | --- |
| ||(Fe→O) | 0 | 0 |
| -||(Fe→O) | 1.99 | 1.80 |

**[3] Relative energies of the two states of [K(1,2-dimethoxyethane)$_4$][(tpa$^{Mes}$)Fe] in which the Fe$^{2+}$ moments are oriented along the ||(Fe→N$_a$) and -||(Fe→ N$_a$) directions**

The crystal structure of [K(1,2-dimethoxyethane)$_4$][(tpa$^{Mes}$)Fe] has four formula units per unit cell. To simplify our calculations for [K(1,2-dimethoxyethane)$_4$][(tpa$^{Mes}$)Fe], we constructed a hypothetical solid that has one [(tpa$^H$)Fe]$^-$ anion and one [K(1,2-dimethoxyethane)$_4$]$^+$ per unit cell with large cell parameters, a = 30 Å, b = 25 Å, c = 27 Å.

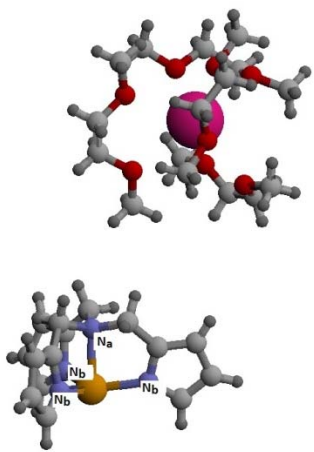

The relative energies ΔE (meV/Fe) of the two states in which the Fe$^{2+}$ ion moments are oriented along the ||(Fe→N$_a$) and -||(Fe→N$_a$) directions were calculated by performing DFT+U+SOC calculations with U$_{eff}$ = 4 and 5 eV, a set of (2×4×2) k-points, the SCF convergence criterion of 10$^{-7}$ eV, and the planewave cutoff energy of 450 eV.

Table S3. Relative energies ΔE (meV/Fe).

|  | U$_{eff}$ = 4 eV | U$_{eff}$ = 5 eV |
| --- | --- | --- |
| ||(Fe→N$_a$) | 0 | 0 |
| -||(Fe→N$_a$) | 6.43 | 6.88 |

## [4] Author Contributions